\documentclass[A4paper,unsortedaddress,superscriptaddress,prb,twocolumn]{revtex4-1}
\usepackage{hyperref}
\usepackage{epsfig}
\usepackage{graphicx}
\usepackage{subfigure}
\usepackage{latexsym}
\usepackage{color}
\usepackage{fullpage}
\usepackage{dcolumn}
\usepackage{bm}
\usepackage{ulem}
\usepackage{units}

\newcommand{\nico}[1]{\textcolor{red}{#1}}

\begin{document}

\title{Intercalating cobalt between graphene and iridium (111): a spatially-dependent kinetics from the edges}

\author{Sergio Vlaic}
\affiliation{CNRS, Inst NEEL, F-38000 Grenoble, France}
\affiliation{Univ. Grenoble Alpes, Inst NEEL, F-38000 Grenoble, France}
\affiliation{LPEM, ESPCI Paris, PSL Research University, CNRS, Sorbonne Universit\'{e}s, UPMC University of Paris 6, 10 rue Vauquelin, Paris F-75005, France}
\author{Nicolas Rougemaille}
\email{nicolas.rougemaille@neel.cnrs.fr}
\affiliation{CNRS, Inst NEEL, F-38000 Grenoble, France}
\affiliation{Univ. Grenoble Alpes, Inst NEEL, F-38000 Grenoble, France}
\author{Amina Kimouche}
\affiliation{CNRS, Inst NEEL, F-38000 Grenoble, France}
\affiliation{Univ. Grenoble Alpes, Inst NEEL, F-38000 Grenoble, France}
\author{Benito Santos Burgos}
\affiliation{Elettra-Sincrotrone Trieste S.C.p.A., S.S: 14 km 163.5 in AREA Science Park, I-34149 Basovizza, Trieste, Italy}
\author{Andrea Locatelli}
\affiliation{Elettra-Sincrotrone Trieste S.C.p.A., S.S: 14 km 163.5 in AREA Science Park, I-34149 Basovizza, Trieste, Italy}
\author{Johann Coraux}
\email{johann.coraux@neel.cnrs.fr}
\affiliation{CNRS, Inst NEEL, F-38000 Grenoble, France}
\affiliation{Univ. Grenoble Alpes, Inst NEEL, F-38000 Grenoble, France}

\begin{abstract}
Using low-energy electron microscopy, we image in real time the intercalation of a cobalt monolayer between graphene and the (111) surface of iridium. Our measurements reveal that the edges of a graphene flake represent an energy barrier to intercalation. Based on a simple description of the growth kinetics, we estimate this energy barrier and find small, but substantial, local variations. These local variations suggest a possible influence of the graphene orientation with respect to its substrate and of the graphene edge termination on the energy value of the barrier height. Besides, our measurements show that intercalated cobalt is energetically more favorable than cobalt on bare iridium, indicating a surfactant role of graphene.
\end{abstract}


\maketitle

\begin{figure}
\begin{center}
\includegraphics[width=7.75cm]{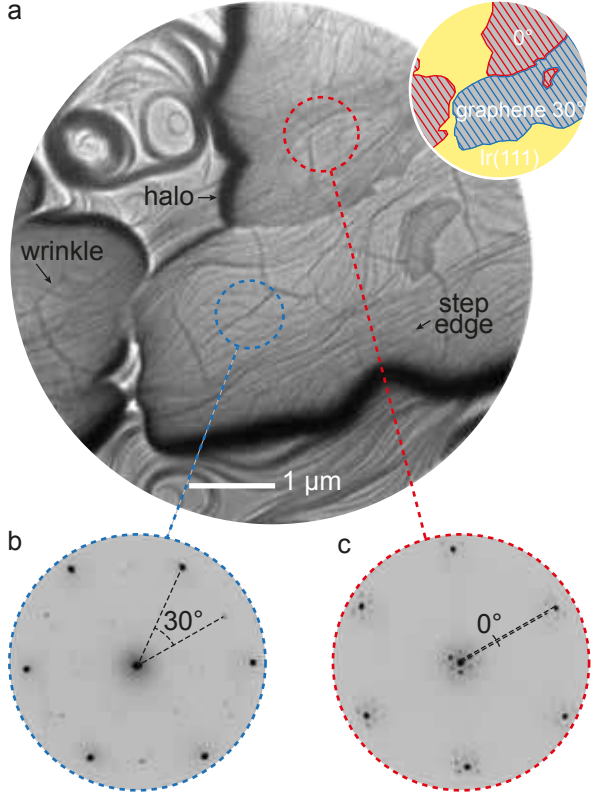}
\caption{\label{fig1}(a) LEEM micrograph (start voltage, 4~V) revealing graphene domains with distinct electron reflectivity. The meandering atomic step edges of the metal surface and the branched network of graphene wrinkles are both visible. The dark halo blurring the graphene edges is a joint effect of the deviation of the electron beam due to the electronic density contrast between graphene and the metal, and of the contrast aperture limiting the angular acceptance of the instrument. Top right: Schematics of the LEEM micrograph permitting identification of graphene-free (yellow) and graphene-covered (gray) regions having different orientations, highlighted in red and blue for 0$^\circ$- and 30$^\circ$-rotated graphene domains, respectively. (b,c) Micro-LEED (start voltage, 40~V) patterns measured at the two locations marked by a dotted colored circle in (a), for a 30$^\circ$- (blue) and 0$^\circ$-rotated (red) graphene domain. The angles between the high symmetry reciprocal space directions of graphene and Ir(111) are indicated.}
\end{center}
\end{figure}

Graphene is an atomically-thin coating that forms at the surface of various carbides and metals.\cite{Oshima,Tetlow} Being impermeable and inert, it protects metal surfaces, which may otherwise lose their properties due to oxidation\cite{Dedkov} or to the adsorption of airborne molecules.\cite{Varykhalov} The continuity and crystalline quality of the graphene layer are essential to protect efficiently the support surface. For instance, the dissociative adsorption of molecular oxygen in the bare regions of the metal may favor intercalation of oxygen below graphene-covered regions, leading to the oxidation of the entire surface.\cite{Prasai,Raman,Dlubak,Kimouche,Ligato,Omiciuolo}

While intercalation might be in some cases regarded as a detrimental process, it also offers a wealth of opportunities to modify graphene's properties. Intercalation is in fact an established route to functionalize graphene from below,\cite{Varykhalov_b,Weser,Calleja,Monazami}, to decouple it from its substrate,\cite{Varykhalov_b,Enderlein,Sutter,Mao,Meng,RomeroMuniz} to modify the properties of intercalated layers,\cite{Decker,Rougemaille,Coraux,Yang,Vu} and to control chemistry underneath graphene.\cite{Sutter,Mu,Drnec,MartinezGalera,Fu} In the case of metal intercalants, several intercalation pathways have been identified involving either pre-existing point defects\cite{Rougemaille,Coraux} or their formation\cite{Sicot,Sicot_b}, and curved regions of graphene, such as wrinkles and substrate step edges.\cite{Vlaic} Identifying and selecting these intercalation pathways is of crucial importance for the preparation of advanced multi-layered functional materials based on high-quality graphene.

Here, we focus on the energetics of an intercalation process taking place at the edges of graphene, which we monitored in real time by means of low-energy electron microscopy (LEEM) observations. In our work, the intercalant is cobalt and graphene is prepared on the (111) surface of iridium. Mass transport through the graphene edge is described using a phenomenological model based on a limited number of free parameters. This allows us to estimate the energy barrier involved in the intercalation mechanism. We find that this energy barrier differs by a few 10~meV typically, from one location to another along the graphene edges. These variations suggest that the nature of the graphene edge and the crystallographic orientation of the graphene flake with respect to the metal substrate have an influence on the rate of intercalation.

\begin{figure*}
\begin{center}
\includegraphics[width=15.11cm]{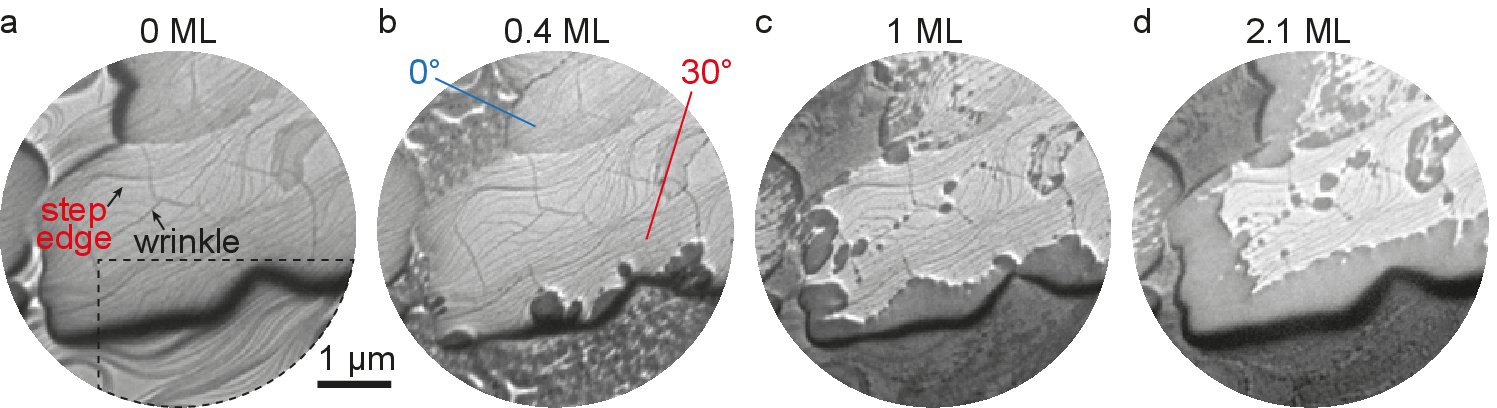}
\caption{\label{fig2a} Series of LEEM (start voltage, 4~V) micrographs after the deposition at 250$^\circ$C of the equivalent of 0 (a), 0.4 (b), 1 (c), and 2.1~ML (d) of Co, for the same region as in Fig.~\ref{fig1}(a). The dotted frame in (a) highlight\nico{s} the region addressed in Fig.~\ref{fig2b}. Arrows indicate a substrate step edge and a wrinkle.\cite{noteonwrinkles}}
\end{center}
\end{figure*}

Graphene was prepared under ultra-high vacuum by catalytic decomposition of ethylene on a clean Ir(111) surface. This metal surface represents a prototypical substrate allowing to grow high-quality, exclusively single layer graphene, with extended lattice continuity and controlled crystallographic orientation.\cite{Coraux_b,Loginova,vanGastel} Ethylene was introduced with a 5$\times$10$^{-8}$~mbar pressure, and the Ir(111) surface was kept at 1000$^\circ$C during graphene growth until about 80\% of the Ir surface was covered. The LEEM images of the surface [Fig.~\ref{fig1}(a)] show graphene domains having different electron reflectivity.\cite{Loginova} Microprobe low-energy electron diffraction ($\mu$-LEED) measurements reveal both a 30$^\circ$ [Fig.~\ref{fig1}(b)] and a 0$^\circ$ [Fig.~\ref{fig1}(c)] orientation of the graphene domains (variants). Cobalt was subsequently deposited using an electron-beam heated evaporator. The sample was kept at 250$^\circ$C during Co deposition (at a rate of 0.08 monolayers (ML) per minute). 

The surface work function changes drastically depending on whether graphene is present or not,\cite{Vlaic} and depending on whether cobalt is on top or below graphene\cite{Rougemaille}. For example, the surface work function of a Co monolayer on Ir is reduced by about 1.7 eV when covered with graphene\cite{Vlaic}. In a low-energy electron microscope such a strong change can be easily monitored \textit{in situ} by measuring the onset of the mirror mode regime. That way, the intercalation process is tracked at video rate.

Figure~\ref{fig2a} shows a sequence of LEEM images during Co deposition, from 0.4 to 2.1 equivalent of a ML. We first note that within the spatial resolution of our microscope, we do not observe the formation of Co clusters on top of the graphene layer, contrary to what is found when Co is deposited at room temperature.\cite{Vlaic} Instead, Co directly intercalates with no need of extra thermal energy (this is measured through the large change of the surface work function). We deduce from this observation that the surface mobility of Co atoms on graphene must be rather high already at 250$^\circ$C. Co atoms then efficiently reach the edges of the graphene flakes and land down on the Ir surface. At this temperature, intercalation is found to occur mainly from the edges of the graphene flakes, while other intercalation pathways\cite{Rougemaille,Coraux,Vlaic} and intercalant/substrate intermixing\cite{Drnec} can be essentially neglected. This can be seen from the LEEM image sequence reported in Fig.~\ref{fig2a}, which reveals that a rim of intercalated material, extending across several hundreds of nanometers, grows as more Co is deposited.

\begin{figure*}[!ht]
\begin{center}
\includegraphics[width=10.56cm]{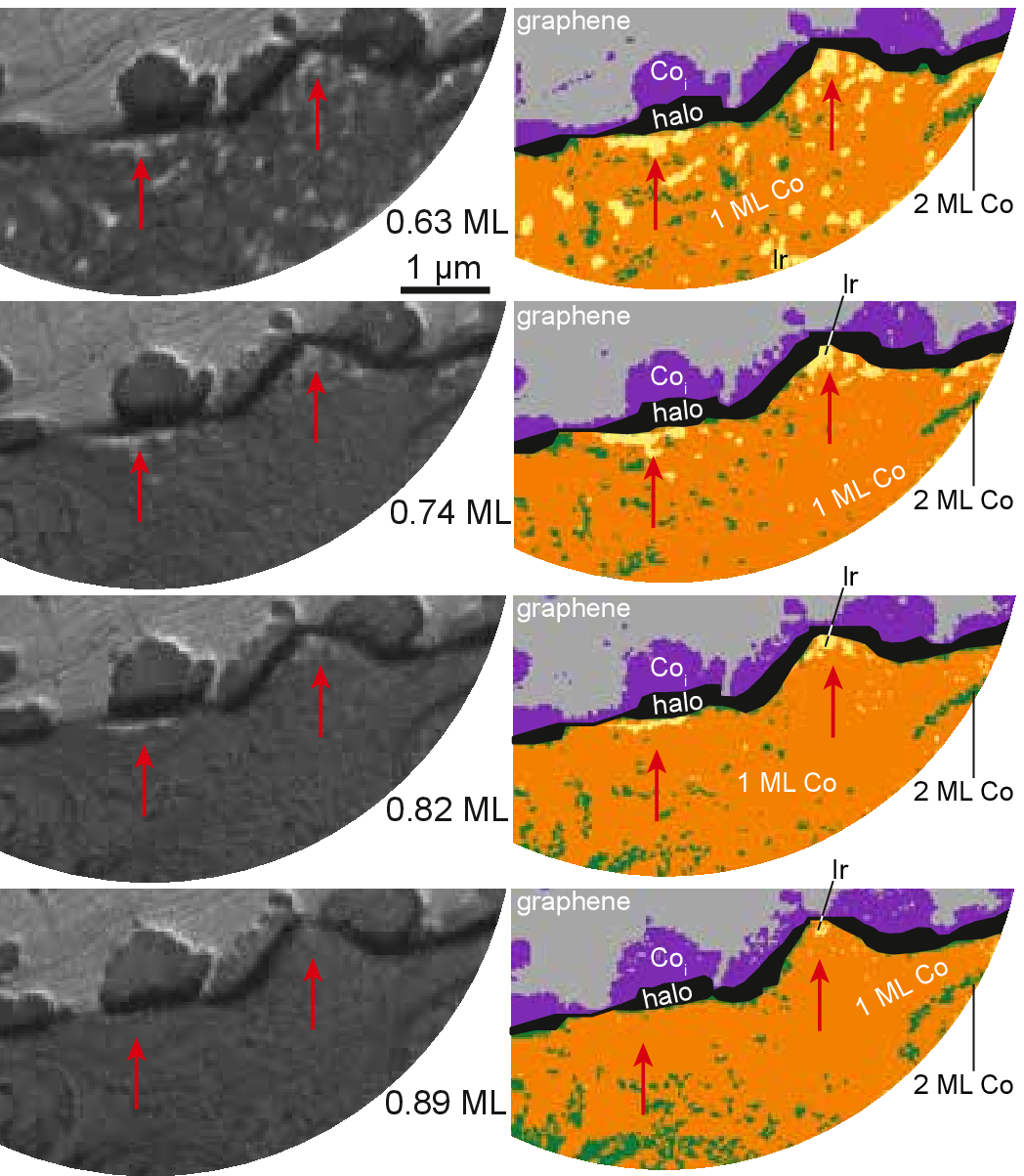}
\caption{\label{fig2b} Series of four LEEM (start voltage, 4~V) micrographs (left side) during Co deposition (the values reported in the figure are expressed in equivalent Co MLs). The area shown corresponds to the dotted frame in Fig.~\ref{fig2a}(a). On the right side, the composition of the surface is highlighted using different color levels, yellow for bare Ir(111), gray for graphene/Ir(111), violet for intercalated Co (Co$_i$), orange for 1~ML Co on Ir(111), and green for 2~ML Co on Ir(111). The black rim represents the halo at the graphene edges [see Fig.~\ref{fig1}(a)]; the red arrows point to Co-free iridium areas close to a graphene edge, on the graphene-free side of the edge. Such areas persist longer close to the graphene edge.}
\end{center}
\end{figure*}

\begin{figure}[!ht]
\begin{center}
\includegraphics[width=6.96cm]{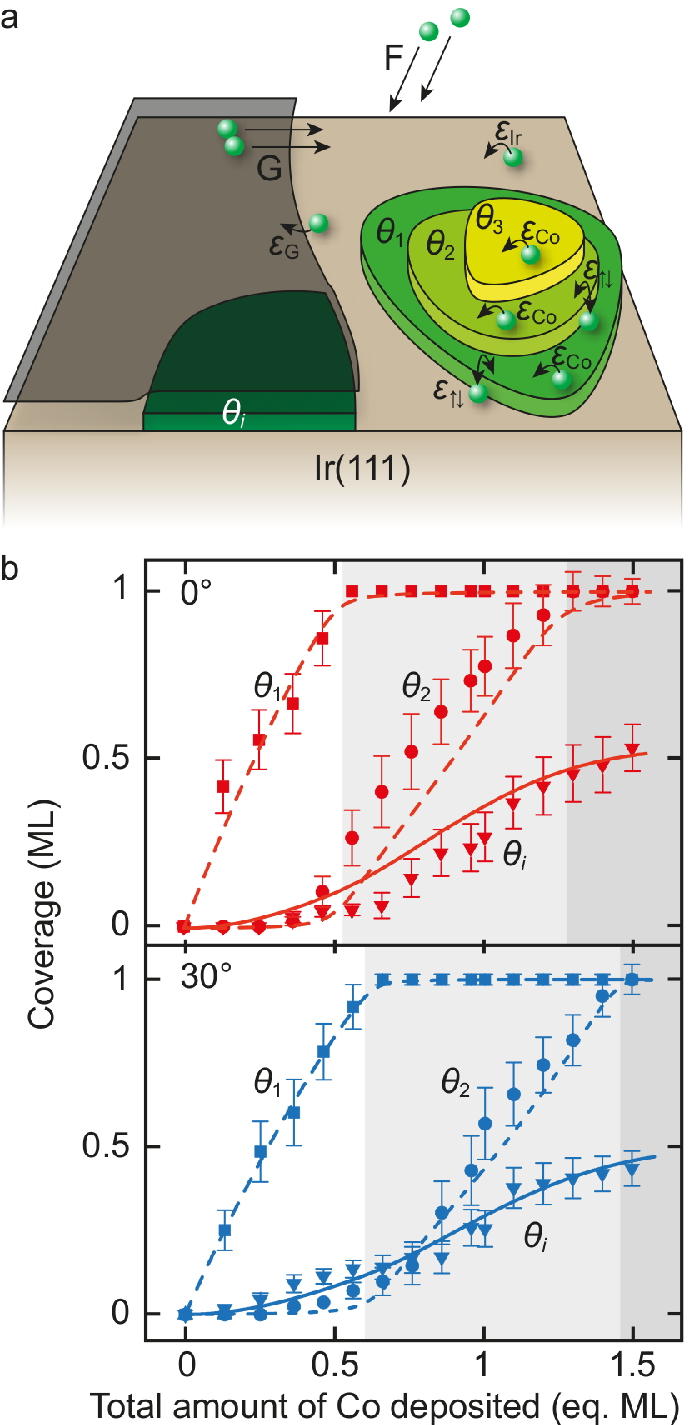}
\caption{\label{fig3} (a) Cobalt coverage deduced from a LEEM movie for $\theta_i$, $\theta_\mathrm{1}$ and $\theta_\mathrm{2}$, as a function of the total amount of Co deposited (for the R0 and R30 variants). Fits to the data using rate equations are shown with dotted and solid lines. The two shades of gray highlight the regions of the graphs where $\theta_\mathrm{1}$ and $\theta_\mathrm{2}$ approach saturation. (b) Schematics of the Ir(111) surface partly covered with graphene, with Co in the form of an intercalated monolayer or in the form of a mono-, bi-, and trilayer on Ir(111) -- with corresponding coverages $\theta_i$, $\theta_\mathrm{1}$, $\theta_\mathrm{2}$, and $\theta_\mathrm{3}$. The processes considered to model the intercalation kinetics are indicated with a curved arrow and activation energies $\varepsilon_\mathrm{G}$, $\varepsilon_\mathrm{Ir}$, $\varepsilon_\mathrm{Co}$, and $\varepsilon_\mathrm{\uparrow}$). The direct impinging flux $F$ of Co atoms (from the evaporator) and the effective additional flux $G$ corresponding to Co atoms first landed on graphene, are also represented. }
\end{center}
\end{figure}

The LEEM image at an early deposition stage (0.4~ML) in Fig.~\ref{fig2a}(b) shows that Co forms mesoscopic 1~ML-thick islands on bare Ir. Below graphene, the Co film is continuous and growth characteristics are different (more step flow like). Careful inspection of the LEEM images on bare Ir, close to the 30$^\circ$-oriented graphene domain, reveals that a few 10/few 100~nm-wide (bright) rim alongside of the graphene edge is Co free [follow \textit{e.g.} the arrows in Fig.~\ref{fig2b}]. We interpret the origin of this Co depleted region as a manifestation of a surface-diffusion-driven mass transfer across the graphene edge. This result is surprising at first sight since the (downward) clamping of graphene edges to the Ir(111) surface\cite{Lacovig,Prezzi,Patera} is expected to hinder intercalation -- an energy barrier $\varepsilon_\mathrm{G}$ [Fig.~\ref{fig3}(a)] needs to be overcome to alter the C-Ir bonding configuration and let Co atoms "pass through". On the basis of this sole argument, Co accumulation, rather than Co depletion, would be expected. 

To account for our observation, an asymmetric energy barrier for Co adatoms passing through, inwards or outwards the graphene edge, needs to be considered, with a higher energy barrier for the outwards path. Two kinds of potential energy landscapes for Co adatom diffusion comply with such an asymmetry. The first kind comprises a dip at the vicinity of the graphene edge, on the graphene-covered side -- where the Co atoms would hence be stabilised. Such a scenario would however contradict a previous report, in which Co atoms were found prominently further inwards from the graphene edges. \cite{Decker} Hence, our data point to a second possible scenario: an asymmetric energy barrier translating a lower average potential landscape on graphene-covered regions than on graphene-free regions -- \textit{i.e.} Co adatoms are thermodynamically more stable when intercalated than on bare Ir(111). The reason for this increased stability might be the higher coordination of the Co atoms below graphene, which form Co-C bonds stronger than Ir-C bonds.\cite{Decker} Overall, graphene can be regarded as a surfactant,\cite{Wang} modifying surface energies. This stabilisation effect is at variance with the case of oxygen adatoms\cite{Granas}: intercalated oxygen is only stable thermodynamically in the presence of an external oxygen pressure, without which oxygen adatoms tend to leave graphene-covered regions and diffuse back towards the bare metal surface.\cite{Granas} 

Figure~\ref{fig3}(a) translates our LEEM observations in a statistical manner, representing, as a function of Co deposition, the surface fraction of Co in the form of an intercalated ML ($\theta_i$) or in the form of a Co ML ($\theta_\mathrm{1}$) and a Co bilayer ($\theta_\mathrm{2}$) on Ir(111). The analysis was performed for two graphene domains, corresponding to 30$^\circ$ and 0$^\circ$ orientations. In both cases, Co growth on bare Ir(111) does not follow a layer-by-layer mode: above the equivalent of 0.4~ML of deposited Co, the second Co layer starts to grow, before the first layer is complete (see in Fig~\ref{fig3}(a) how $\theta_\mathrm{2}$ increases before $\theta_\mathrm{1}$ saturates). Besides, the efficiency of Co intercalation appears enhanced when the second Co layer forms on Ir (see in Fig.~\ref{fig3}(b) how $\theta_\mathrm{i}$ increases together with $\theta_\mathrm{2}$), and is already significant before $\theta_\mathrm{1}$ saturates. This points to a non-negligible energy barrier $\varepsilon_\mathrm{G}$ for Co atoms to intercalate, as mentioned above, but also to an energy barrier $\varepsilon_\mathrm{\uparrow}$ for Co atoms to climb up Co atomic step edges.

A key observation is that Co intercalation progresses differently at the vicinity of two graphene domains, as can be seen in Fig~\ref{fig3}(a) ($\theta_\mathrm{i}$ increases differently for both graphene domains). In particular, Co intercalation sets in earlier for the 30$^\circ$-oriented domain (see bottom panel in Fig.~\ref{fig3}(a) in the range of 0.25 to 0.7 equivalent ML of Co deposited). The effect is also visible in Fig.~\ref{fig2a}(b) for 0.4 equivalent ML deposited on the surface. This is consistent with the fact that the completion of the first and second Co ML on Ir ($\theta_\mathrm{1}$ and $\theta_\mathrm{2}$) are both delayed for this domain compared to the 0$^\circ$ one [see the shifted gray and dark areas in Fig~\ref{fig3}(a)]. These experimental findings suggest that the energy barrier $\varepsilon_\mathrm{G}$ for Co atoms to intercalate is different in the two cases, smaller for the 30$^\circ$ graphene domain, where intercalation is more efficient.

To gain more quantitative insights, we describe the time-evolution of $\theta_i$, $\theta_\mathrm{1}$, $\theta_\mathrm{2}$, and $\theta_\mathrm{3}$ (3~ML of Co on Ir) using a simple material balance model. All kinetic processes we consider are represented in Fig~\ref{fig3}(b). They include the incoming Co flux, $F$, (from the Co evaporator) and the local surface Co flux, $G$, originating from the fast diffusion of Co atoms atop the graphene flake (no formation of Co clusters is observed on graphene). Mass transport on Ir(111) and atop Co layers is accounted for by the energy barriers $\varepsilon_\mathrm{Ir}$ and $\varepsilon_\mathrm{Co}$ respectively. Similarly, mass transport hinderance when climbing up atomic step edges on graphene-free regions is accounted for by an energy barrier $\varepsilon_\mathrm{\uparrow}$, while mass transport across graphene edges, towards graphene-covered regions, is accounted for by an energy barrier $\varepsilon_\mathrm{G}$ implying (at least partly) the breaking of metal-carbon bonds (we assume that $\varepsilon_\mathrm{G}$ at a given location is constant all along the intercalation process). We simplify the asymmetric energy barrier (see discussion above) for Co to enter/escape graphene-covered regions, by assuming an infinite barrier for escaping. A set of four coupled differential equations can then be written (more details can be found in the supplementary material) that we solve numerically. The best fit to the data is shown in Fig.~\ref{fig3}(a), and corresponds to $\varepsilon_\mathrm{Co}$ = 0.14~eV (while $\varepsilon_\mathrm{Ir}$ is set to 0.2~eV), $\varepsilon_\mathrm{\uparrow}$ = 0.20~eV, and $\varepsilon_\mathrm{G}$ = 1.46 and 1.48~eV for the two different graphene domains. Although the latter difference of 20~meV is small, it is significant.

To explain this difference several effects can be invoked. One is the orientation of the substrate step edges with respect to the edges of the graphene flakes: for example, perpendicular substrate step edges could facilitate Co diffusion and be thought as tunnels for Co intercalation. Another effect could be related to the graphene edge orientation, which is expected to modify the strength of the graphene edge-substrate interaction, and thus its permeability to Co (and in turn, $\varepsilon_\mathrm{G}$). A last effect is the different interaction between graphene and Ir(111) depending on the graphene domain orientation,\cite{Starodub} which is liable to control the thermodynamic surfactant effect of graphene: Co atoms should be stabilised in a different manner whether they are below a 30$^\circ$ or a 0$^\circ$ rotational domain. Discriminating these three effects is beyond the scope of the present work. More information about the chemistry of the graphene-Ir and graphene-Co interactions, including interactions at graphene edges, would presumably be insightful to this end.

In summary, we find that Co intercalation between graphene and Ir(111), from the graphene edges, can be activated at 250$^\circ$C. A Co depletion alongside the graphene edges, on graphene-free regions, signals an asymmetric energy barrier to Co intercalation, and a thermodynamically more favorable binding configuration underneath graphene than on bare Ir(111). This suggests a surfactant role of the graphene layer, which is expectedly domain orientation-dependent. Beyond these thermodynamics considerations, kinetic hinderance play an important role. In particular, we infer substantial energy barriers to intercalation, and observe that they vary spatially. We ascribe this spatial dependency to the crystallographic orientation of graphene with respect to its substrate, to the configuration of the carbon-metal bonds at graphene edges, and/or to the contribution of the substrate step edges. Based on a crude model, the variations in energy barriers are small but significant, estimated to be of the order of a few 10 meV.

We emphasize that intercalation has been demonstrated for a variety of elements, including various transition metals and alkali atoms, between graphene and a variety of substrates. The role of the interactions between the intercalated atoms and the substrate atoms might be an important factor governing the intercalation process. Strain in the intercalated layer (compared to the same material in a bulk phase) and polarisation of the interatomic bonds at the interface between the intercalated layer and the substrate, might also influence the intercalant diffusion below graphene. We hence expect qualitatively similar intercalation processes for Fe, Co, and Ni atoms, and for Ir(111) and Pt(111) substrates.

The authors thank Laurence Magaud, Benjamin Canals, and Tevfik Onur Mente\c{s} for fruitful discussions. S.V. acknowledges support by the Swiss National Science Foundation through project PBELP2-146587 and J.C. acknowledges support from the French Agence Nationale pour la Recherche contracts ANR-2010- BLAN-1019-NMGEM, ANR-12-BS-1000-401-NANOCELLS and the European Community EU-NMP3-SL-2010-246073 GRENADA contract.


\begin{thebibliography}{41}
\bibitem [{not()}]{noteonwrinkles}%
  \BibitemOpen
  \href@noop {} {}\bibinfo {note} {Graphene wrinkles and substrate step edges
  show different contrast in LEEM, especially when adjusting the focus of the
  microscope. The wrinkles are also identified as they form upon cool down,
  following the growth of graphene on the metal
  surface.\cite{Loginova,Ndiaye,Sutter_b}}\BibitemShut {NoStop}%
\end{thebibliography}

%

\end{document}